\documentclass{PoS}
\usepackage{amsmath,amssymb,amsfonts}
\usepackage{slashed}
\usepackage{bbm}
\usepackage{cite}
\usepackage[utf8]{inputenc}
\usepackage[english]{babel}

\title{Finetuning the continuum limit in low-dimensional supersymmetric theories
\footnote{We thank Raphael Flore and Christian Wozar for their active
collaboration and Georg Bergner and Mithat Ünsal for helpful discussions.}}
\ShortTitle{Finetuning the continuum limit in low-dimensional supersymmetric theories}
\author{Bjoern H. Wellegehausen\\%
        Justus-Liebig-University Giessen\\
        E-mail: \email{bjoern.wellegehausen@theo.physik.uni-giessen.de}}
\author{\speaker{Daniel Koerner} and Andreas Wipf\\%
        Friedrich-Schiller-University Jena\\
        E-mail: \email{daniel.koerner@uni-jena.de}, \email{wipf@tpi.uni-jena.de}}

\abstract{Supersymmetry is a prominent candidate for physics beyond the 
standard model. In order to compute the spectrum of supersymmetric theories, 
we employ nonperturbative lattice QFT techniques which due 
to the discretisation of spacetime violate supersymmetry at finite lattice 
spacings. Care has to be taken then to restore supersymmetry in the continuum 
limit. We discuss a discretisation of the supersymmetric Nonlinear O(N) Sigma 
model in two dimensions and argue that supersymmetry may be restored by 
finetuning of a single parameter. Furthermore, we show preliminary results 
for the vacuum physics of 
$\mathcal{N}=2$ Super-Yang-Mills theory in three dimensions.}

\FullConference{31st International Symposium on Lattice Field Theory LATTICE 2013\\
		 July 29 - August 3, 2013\\
		 Mainz, Germany}

\graphicspath{{./img/}}

\DeclareMathOperator{\tr}{tr}
\newcommand{\komm}[2]{\ensuremath{\left[\,#1,#2\,\right]}}

\newcommand{\ii}{\mathrm{i}}
\newcommand{\erw}[1]{\left \langle #1 \right \rangle}
\newcommand{\bo}[1]{\boldsymbol{ #1 }}
\newcommand{\m}[1]{\mathbf{ #1 }}
\newcommand{\cN}{\mathcal{N}}

\newcommand{\Neqtwo}{\cN\!\!=\!2}

\newcommand{\includeEPSTEX}[1]{\includegraphics{#1}}

\begin{document}
\section{Beyond the Standard Model - Supersymmetry}
\noindent The Standard Model provides a very successful description of electroweak 
and strong interactions. However, cosmological observations uncover 
discrepancies from the theoretical prediction, i.e. the matter to 
antimatter ratio or dark matter. These observations point to physics 
beyond the Standard Model. Supersymmetry, an extension of the Poincaré 
symmetry of spacetime transformations, provides a possible extension 
of the Standard model. The algebra of supercharges closes on the
energy-momentum,
\begin{equation}
\{Q_{\alpha},\bar{Q}_{\beta}\} = 2 \gamma^{\mu}_{\alpha\beta} P_\mu . 
\label{eq:algebra}
\end{equation}
On a discrete spacetime lattice, infinitesimal translations or rotations
are not possible and hence supersymmetry is inevitably broken at finite 
lattice spacing. In order to restore supersymmetry in the continuum 
limit, all relevant susy-breaking operators need to be controlled. A 
recent approach \cite{catterall09} tries to implement a part of the supersymmetry exactly
on the lattice, which may guarantee the correct continuum limit. However,
we will show that this approach fails for the supersymmetric nonlinear 
Sigma Model and we then proceed to argue that a finetuning of 
susy-breaking operators leads to the expected continuum limit. 
Furthermore we show that for $\mathcal{N}=2$ supersymmetric Yang-Mills theory
a supersymmetric continuum limit is possible by finetuning a single parameter.

\section{Supersymmetric O(3) nonlinear $\sigma$ model}
\noindent The supersymmetric extension of the bosonic O(3) nonlinear 
Sigma model is formulated most elegantly in superspace, where the 
usual constraint is applied to the superfield. To obtain the action in 
regular spacetime, we expand the superfield and integrate out the 
auxiliary field with the result
\begin{equation}
S = \tfrac{1}{2g^2} \int d^2 x ~ \left(\partial^{\mu}\mathbf{n} \partial_{\mu}\mathbf{n} + i\bar{\bo{\psi}}\slashed{\partial}\bo{\psi} + \tfrac{1}{4}(\bar{\bo{\psi}} \bo{\psi})^2\right), \quad\text{where }\m{n}^2 = 1, \quad\m{n}\bo{\psi} = 0.
\end{equation}
In order to simulate the theory on the lattice, it is necessary to 
rewrite the contraints for the component fields. We choose a 
stereographic projection of the superfield which resolves the
field constraints explicitly \cite{sigmapaper}. In addition to supersymmetry, 
the classical action is invariant under SO(3)-rotation of the fields.
Furthermore, a discrete $\mathbb{Z}_2$ chiral symmetry is broken 
spontaneously on the lattice for arbitrary coupling and is connected to
the dynamical generation of mass. \\
Supersymmetry relates the bosonic and fermionic components of the superfields,
\begin{equation}
\delta_I n_i = i\bar{\epsilon}\psi_i,\quad\delta_I \psi_i^{\alpha} = \slashed{\partial} n_i \epsilon + \tfrac{i}{2} (\bar{\psi}\psi) n_i\epsilon^{\alpha}.
\label{eq:firstsusy}
\end{equation}
Since the target space $S^2$ is Kähler, there exists an $\Neqtwo$-supersymmetric 
extension of the model. The second susy transformation reads
\begin{equation}
\delta_{II}\bo{n} = ~\bo{n}\times i\bar{\epsilon}\bo{\psi},
\quad\delta_{II} \bo{\psi}^{\alpha} = - \bo{n} \times (\slashed{\partial} \bo{n} \epsilon)^{\alpha} - i\bar{\epsilon}\bo{\psi} \times \bo{\psi}^{\alpha}.
\label{eq:secondsusy}
\end{equation}	
The supersymmetries are generated by the supercharges:
\begin{equation}
 Q^{I}_{\alpha} = \int i \partial_{\mu} n_i \gamma^{\mu} \gamma^0 \psi_i,\quad Q^{II}_{\alpha} = \int -i \epsilon_{ijk} n_i \partial_{\mu} n_j \gamma^{\mu} \gamma^0 \psi_k.
\end{equation}

\subsection{Implementing exact supersymmetry and O(3) symmetry - a no-go theorem}
\begin{figure}
\parbox{0.99\columnwidth}{
\begin{center}
\resizebox{0.31\columnwidth}{!}{\input{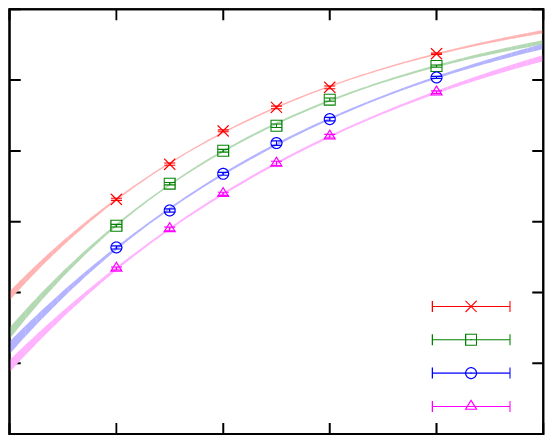}}\hspace{10px}
\resizebox{0.29\columnwidth}{!}{\includeEPSTEX{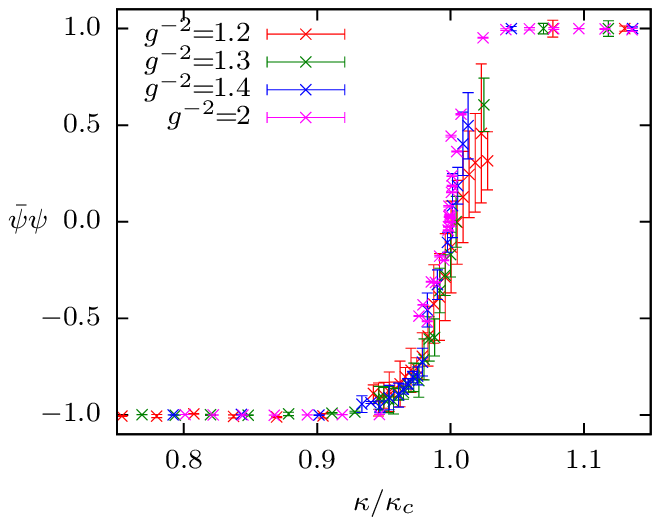}}\hspace{10px}
\resizebox{0.31\columnwidth}{!}{\includeEPSTEX{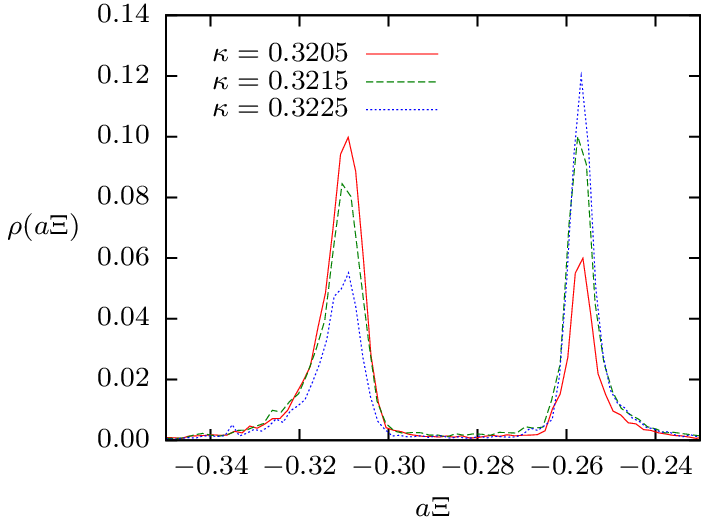}}\\
\end{center}
}
\caption{\textbf{Left panel:} measurement of the order parameter $\left< n_3 \right>$
in the $Q$-exact ensemble for different coupling constants and lattice volumes.
\textbf{Center panel:} phase transition of the renormalized chiral 
condensate at the critical finetuning parameter $\kappa_c$. 
\textbf{Right panel:} $\mathbb{Z}_{2}$-symmetric histogram of the 
chiral condensate.}
\label{fig:brokeno3}
\end{figure}

\noindent The authors of \cite{catterall06} construct a 
nilpotent supercharge $Q^2 = 0$, such that the action may be written as
$S = Q  \Lambda$. By implementing this relation exactly on the lattice,
part of the supersymmetry is respected and the supersymmetric continuum 
limit may be approached without the need for finetuning. However, this "$Q$-exact 
action" breaks the O(3) symmetry explicitly at any lattice spacing. 
Mermin-Wagner's theorem dictates that the O(3) symmetry cannot be broken 
spontaneously and one therefore needs to take care in restoring
the symmetry in the continuum limit. Alas, by simulating the action of
\cite{catterall06} we find that the order parameter of the O(3) symmetry
does not vanish in the infinite volume limit, as is shown in Fig. 
\ref{fig:brokeno3} (left panel), and the resulting continuum limit does not belong 
to the supersymmetric O(3) NLSM, therefore rendering the $Q$-exact 
discretisation invalid \cite{sigmapaper}.
This failure raises the question whether it is possible at all to find a
lattice formulation of this model that preserves both the O(3) symmetry
as well as part of supersymmetry. We will argue that this is indeed
not possible. A general symmetry of the model needs to leave invariant
both the action as well as the constraints. The first susy transformation
in eq. \ref{eq:firstsusy} breaks the constraint $n\psi = 0$ at finite
lattice spacing,
\begin{equation}
\delta_I(n_x\psi_x^{\alpha}) = i\bar{\epsilon}\psi_x \psi_x^{\alpha} + \sum_{y\in L} n_x D_{xy}^{\alpha \beta} n_y \epsilon^{\beta}  + \tfrac{i}{2} (\bar{\psi}_x \psi_x) n^2_x \epsilon^{\alpha} = \sum_{y\in L} n_x D_{xy}^{\alpha \beta} n_y \epsilon^{\beta}.
\end{equation}
The second susy transformation always respects the constraints
by use of vector identities,
\begin{equation}
\delta_{II}(n_x \psi_x) = \sum_{y\in L} n_x \cdot (n_x \times D_{xy}^{\alpha \beta} n_y \epsilon^{\beta}) = 0,\quad \delta_{II}(n^2_x) = 2 n_x (n_x \times i\bar{\epsilon}\psi_x) = 0.
\end{equation}
However, the second susy transformation by itself cannot be a symmetry on the lattice
since the susy algebra cannot be closed (eq. \ref{eq:algebra}) and
any nontrivial combination of the susy transformations $\delta_I$ and
$\delta_{II}$ is not compatible with the constraints. 
The approach in \cite{catterall06} contains an additional topological 
charge, which however does not alter the supersymmetry transformations
and thus the problem persists. One may further introduce nonlocal interaction
terms like $\sum_{y,z,w} C_{xyzw}\psi_x\psi_y\psi_z\psi_w$ instead of 
$(\psi_x)^4$. But none of such terms is able to cure the problem at hand. 
The last remaining possibility is the introduction of terms that vanish 
in the continuum limit, but render the lattice action invariant under 
$Q^{II}$ for finite lattice spacing. Once again, no appropriate terms
are available. We thus conclude that no discretization of the O(3)-NLSM 
exists which maintains O(3)-invariance and exact supersymmetry.

\subsection{Finetuning the supersymmetric continuum limit}
\begin{figure}
\parbox{0.99\columnwidth}{
\begin{center}
\resizebox{0.28\columnwidth}{!}{\includeEPSTEX{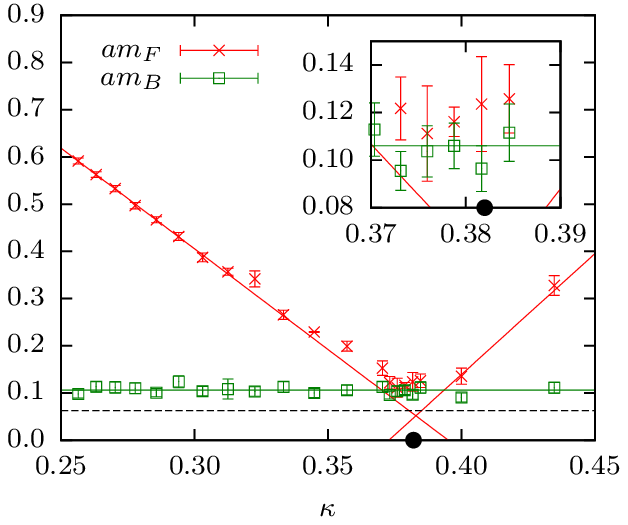}}\hspace{10px}
\resizebox{0.29\columnwidth}{!}{\includeEPSTEX{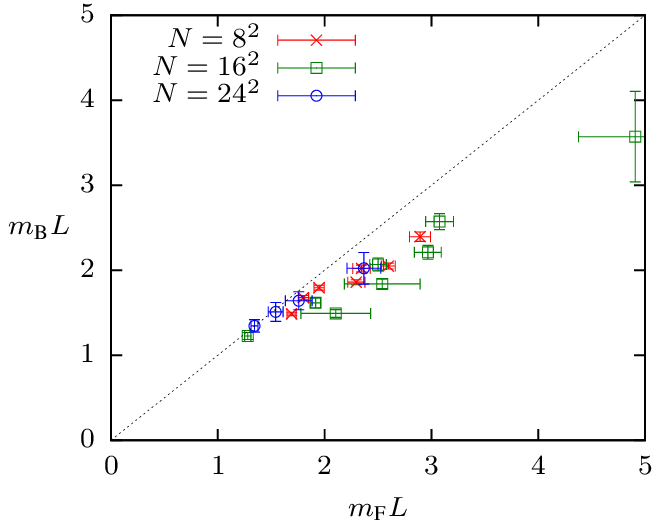}}\hspace{10px}
\resizebox{0.32\columnwidth}{!}{\includeEPSTEX{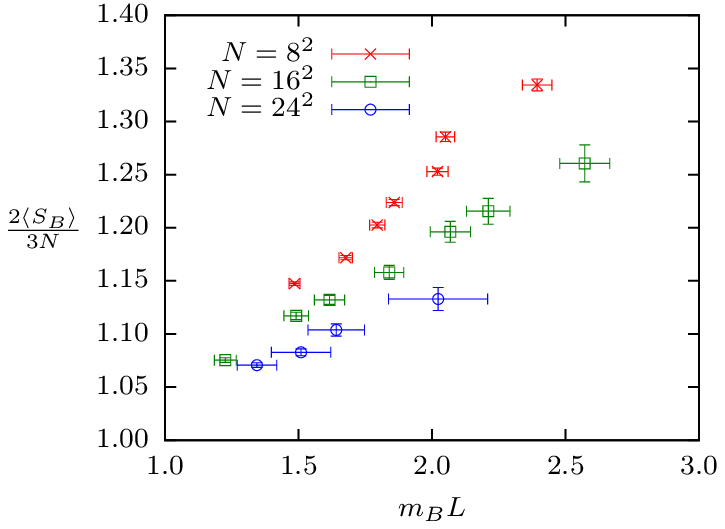}}
\end{center}
}
\caption{\textbf{Left panel:} finetuning $\kappa$ 
towards the critical value $\kappa_c$ reveals a possible mass degeneracy. 
\textbf{Center panel:} bosonic and fermionic mass gap for Wilson 
fermions at $\kappa=\kappa_{c}$.
\textbf{Right panel:} keeping the box size $m_BL$ fixed, the simple Ward
identity is restored in the continuum limit.}
\label{fig:wilsonmasses}
\end{figure}

\noindent In order to discretize the fermionic degrees of freedom, we 
replace the continuum Dirac operator by the Wilson lattice operator
which suppresses unwanted fermion doublers. While being ultralocal, the
Wilson operator introduces an explicit breaking of chiral symmetry 
which should be cancelled by finetuning the hopping parameter $\kappa$.
In figure \ref{fig:brokeno3} (center panel) we show the first order phase transition 
of the renormalized chiral condensate which marks the critical value 
$\kappa_c$. At $\kappa=\kappa_c$, we can cleary distinguish the two minima of the
chiral condensate in the right panel of figure \ref{fig:brokeno3}.
Supersymmetry predicts that the masses of the elementary bosonic 
and fermionic excitations coincide in the continuum limit.
Adjusting the hopping parameter as in figure \ref{fig:wilsonmasses} (left panel), 
we see that the bosonic mass stays constant when we vary $\kappa$, 
whereas the fermionic mass 
shows a linear dependence in a large range. In the vicinity of $\kappa_c$, this linear 
dependence ceases and the fermionic mass does not go to zero but
stays at the level of the bosonic mass, hinting at a possible degeneracy.
Using the finetuned ensemble in figure \ref{fig:wilsonmasses} (center panel) we
find a residual discrepancy of the masses for intermediate box sizes $m_FL$, 
which can however be attributed to a thermal contribution to the mass gap. 
A similar behaviour can be seen in the Wess-Zumino-model \cite{kastner08}
and we expect it to vanish
for large box sizes. The bosonic action constitutes a simple Ward identity for susy 
restoration, $\left< S_{B} \right> =  \frac{3}{2}N$, and we find that
the relation is restored in the continuum limit (see right panel 
of Fig. \ref{fig:wilsonmasses}). Both the observed masses and the Ward 
identity show a restoration of supersymmetry when $\kappa$ is finetuned.

\section{$\mathcal{N}=1$ supersymmetric Yang-Mills theory in 4 dimensions}
\noindent
The on-shell action in four spacetime dimensions is given by
\cite{Ferrara1974239,Veneziano1982231}
\begin{equation}
S_\mathsf{SYM}=\int d^4 x\,\tr\left\lbrace-\frac{1}{4}F_{\mu\nu}
F^{\mu\nu}+\frac{i}{2}\bar{\lambda} \gamma_\mu
D^\mu\lambda + m\,\bar{\lambda} \lambda \right
\rbrace,\label{eq:D4SUSYAction}
\end{equation}
where $F_{\mu\nu}=\partial_\mu A_\nu-\partial_\nu A_\mu-i g\komm{A_\mu}{A_\nu}$ is the usual field strength tensor
with dimensionless gauge coupling constant $g$.
The gauge field $A_\mu$ and the Majorana spinor $\lambda$ transforms under the adjoint representation of
the gauge group $SU(N_\mathsf{c})$. For $m=0$ the action \eqref{eq:D4SUSYAction} is
invariant under $\mathcal{N}=1$ supersymmetry transformations
\begin{equation}
\delta A_\mu=\ii\, \bar{\epsilon}\, \gamma_\mu \lambda \quad,\quad
\delta \lambda=\ii \, \Sigma_{\mu\nu} F^{\mu\nu} \epsilon \quad \text{and} \quad
\delta \bar{\lambda}=-\ii \, \bar{\epsilon}\,\Sigma_{\mu\nu} F^{\mu\nu},
\end{equation}
where $\epsilon$ is an arbitrary constant Majorana spinor. The gluino mass term $m$ introduces a soft breaking of supersymmetry. The action \eqref{eq:D4SUSYAction} is
invariant under a chiral
$U(1)_\mathsf{A}$ transformation (R-symmetry) that is broken by the chiral anomaly to the discrete subgroup
$\mathbbm{Z}(2N_\mathsf{c})$. A non-vanishing gluino condensate
$\erw{\bar{\lambda} \lambda}\neq 0$ breaks the discrete symmetry further down to 
$\mathbbm{Z}(N_\mathsf{c})$, leaving $N_\mathsf{c}$ inequivalent ground states of
the theory. For $N_\mathsf{c}=2$, two
degenerate ground states should exist, that can be distinguished by the sign of the
gluino condensate. It is expected that the
theory possesses a first order phase transition at vanishing renormalized gluino
mass, if the chiral symmetry is spontaneously broken \cite{Kirchner:1998mp}.
In \cite{Veneziano1982231}, Veneziano and Yankielowicz argued, that the only
supersymmetry breaking operator is related to a non-vanishing gluino condensate. In order to restore
supersymmetry on the lattice, it is therefore sufficient to fine-tune the theory
to a massless gluino in the continuum limit. Due to confinement, the gluino is not
part of the physical spectrum, and it is not possible to measure its mass
directly. But the OZI rule (known from QCD) relates the renormalized gluino mass to the pion mass as $m_g \propto m_\pi^2$ \cite{Donini:1997hh}. 
Very recently, $\mathcal{N}=1$ SYM theory in four
dimensions has been investigated with much effort on the lattice \cite{Curci:1986sm,Donini:1997hh,Farchioni:2004fy,Bergner:2011wf,Giedt:2008xm}. 
But so far the results for the mass spectrum are not conclusive.

As in Yang-Mills theories, it is believed that in SYM
theories only colourless asymptotic states exist and a mass gap is
dynamically generated. Veneziano and Yankielowicz (VY) \cite{Veneziano1982231} and later Farrar-Gabadadze-Schwetz (FGS) \cite{Farrar:1997fn} proposed an effective
Lagrangian, that leads to the particle spectrum shown in Table \ref{tab:VYMultiplett}.
\begin{table}[htb]
\begin{center}
\begin{tabular}{c|lccccc}
multiplet & particle & operator  & spin & mass & SYM-name & QCD-name \\
\hline\hline & $1$ pseudoscalar boson & $\tr \,\bar{\lambda}\gamma_5\lambda$ & $0$ &
$m_{\tilde{g}\tilde{g}}^{0-}$ & $a-\eta^\prime$ & $\eta^\prime$ \\
VY & $1$ scalar boson & $\tr \,\bar{\lambda}\lambda$ & $0$ &
$m_{\tilde{g}\tilde{g}}^{0+}$ & $a-f_0$ & $f_0$ \\
& $1$ Majorana fermion & $\tr \, F_{\mu\nu}\Sigma^{\mu\nu}\lambda$ & $\frac{1}{2}$ &
$m_{g\tilde{g}}$ & gluino-glueball & - \\ \hline
& $1$ scalar boson & $\tr \, F^{\mu\nu}F_{\mu\nu}$ & $0$ &
$m_{gg}^{0+}$ & $0^+$ - glueball & $0^+$ - glueball\\
FGS & $1$ pseudoscalar boson &
$\tr \, F^{\mu\nu}\tilde{F}_{\mu\nu}$ & $0$ & $m_{gg}^{0-}$ &
$0^-$ - glueball & $0^-$ - glueball \\ 
& $1$ Majorana fermion & $\tr \, F_{\mu\nu}\Sigma^{\mu\nu}\lambda$
& $\frac{1}{2}$ & $m_{g\tilde{g}}$ & gluino - glueball & -
\end{tabular}
\caption{Particles of the Veneziano-Yankielowicz (VY) and
Farrar-Gabadadze-Schwetz (FGS) multiplet.}
\label{tab:VYMultiplett}
\end{center}
\end{table}
If supersymmetry is broken by a gluino mass term, the
masses inside one multiplet are no longer degenerate and the $0^+$ glueball
should be the lightest particle.

\section{$\mathcal{N}=2$ supersymmetric Yang-Mills theory in 3 dimensions}

\begin{figure}[htb]
\begin{center}
\scalebox{0.55}{\includeEPSTEX{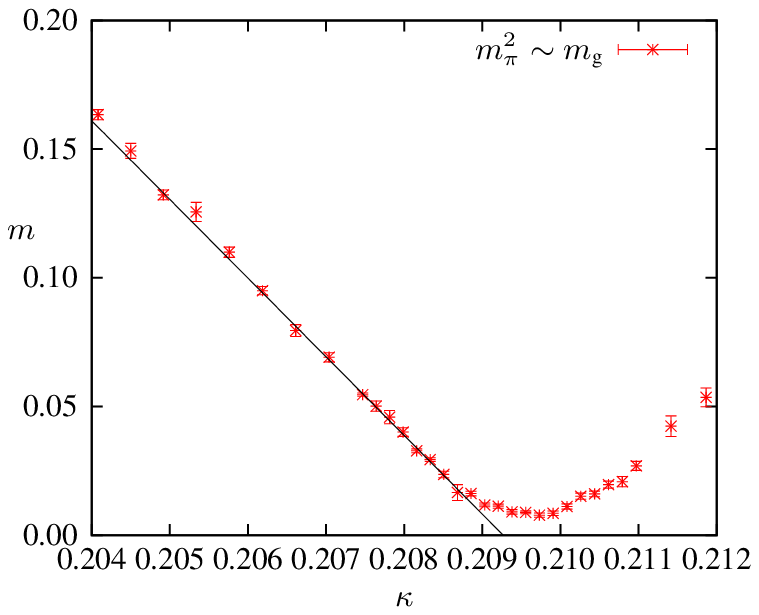}}\hskip5mm
\scalebox{0.55}{\includeEPSTEX{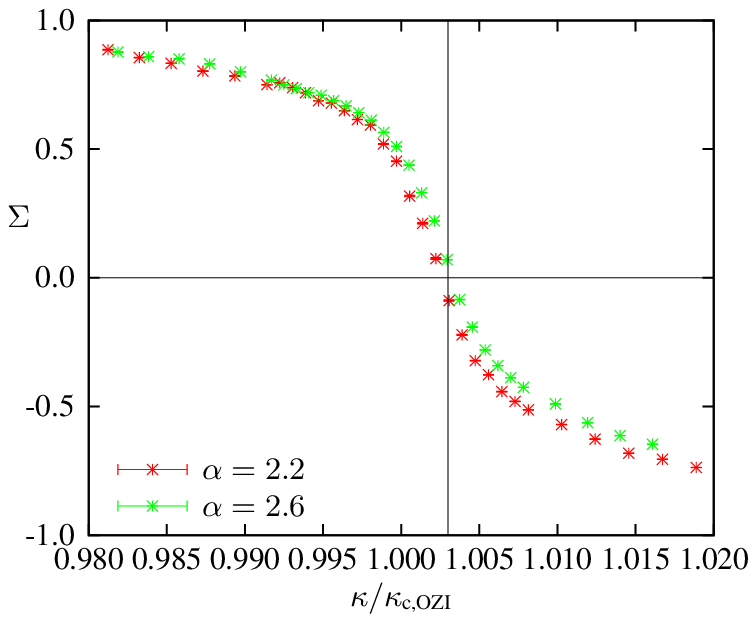}}\hskip5mm
\scalebox{0.55}{\includeEPSTEX{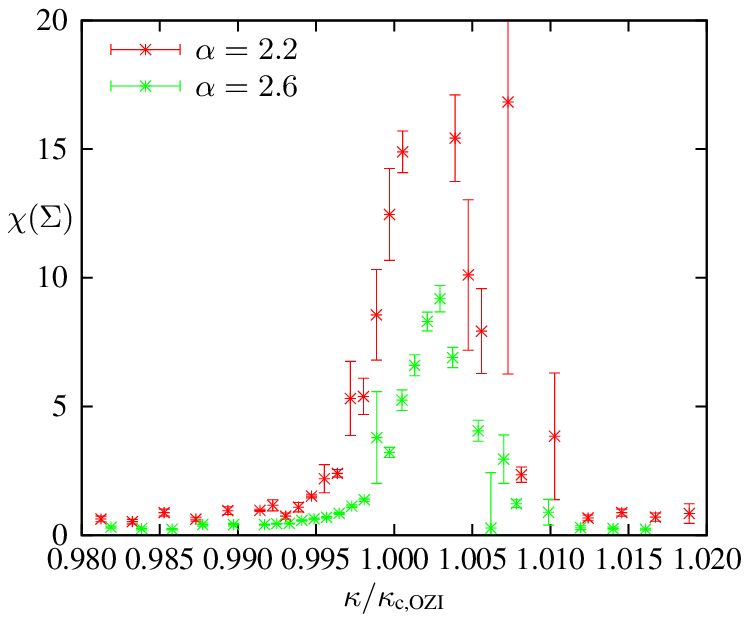}}
\caption{Gluino mass (square of the pion mass) on a $16^2 \times 32$
lattice for $\alpha=2.2$ (left panel). The black 
lines represent a linear fit to the limit of a vanishing gluino mass. The Renormalized chiral condensate (center panel) and the chiral susceptibility
(right panel) are shown for $\alpha=2.2$ and $\alpha=2.6$.}
\label{fig:gluinoMasskappac}
\end{center}
\end{figure}

\noindent Here we investigate the theory in a dimensional reduced version in three dimensions. The action for this (euclidean) $\mathcal{N}=2$ supersymmetric Yang-Mills theory is then given by\footnote{We consider only a gluino mass term to fine-tune the theory, since we expect a scalar mass term to be small in the continuum limit. Nevertheless we have to compare our results to the effective scalar potential in \cite{Unsal:2007jx}.}
\begin{equation}
\begin{aligned}
S^\text{E}=& \alpha \int d^{3}x\, \tr \left \lbrace
\frac{1}{4}F_{IJ}F_{IJ}+\frac{1}{2} \bar{\lambda} \gamma_I^\text{E} D_I \lambda
+\frac{1}{2} D_I \phi D_I \phi - \frac{1}{2}
\bar{\lambda} \gamma_3^\text{E} \komm{\phi}{\lambda} + m \bar{\lambda}
\lambda\right \rbrace.
\end{aligned}
\end{equation}
On a $16^2 \times 32$ lattice for Wilson fermions, for different values of the overall gauge
coupling $\alpha$, the critical hopping parameter $\kappa_\text{c,OZI}(\alpha)$ is
determined such that the gluino becomes massless. In
Fig.~\ref{fig:gluinoMasskappac} (left panel) the square of the pion
mass is shown for $\alpha=2.2$.
A linear fit to the gluino mass yields $\kappa_\text{c,OZI}$ in Tab. \ref{tab:kappac}. Due to the residual Wilson mass at a
finite lattice spacing, the chiral condensate is renormalized as
$\Sigma_\text{ren}(\alpha,\kappa)=Z_1\left(\Sigma(\alpha,\kappa)-m_\text{res}
\right)=Z_1 \Sigma(\alpha,\kappa)-Z_2 \kappa-Z_3$,
where it is assumed that the residual Wilson mass is a linear function in
$\kappa$ (as it is the gluino mass). The renormalization constants $Z_1, Z_2$
and $Z_3$ are fixed such that
$\Sigma_\text{ren}(\alpha,\kappa\ll\kappa_c)=1=-\Sigma_\text{ren}(\alpha,\kappa\gg\kappa_c)$.
The critical point obtained from the vanishing of the pion mass can be compared
to the critical point obtained from the transition in the chiral condensate. 
Therefore in the center and right panel of Fig.~\ref{fig:gluinoMasskappac}
the chiral condensate and its susceptibility are shown as a
function of $\kappa$, normalized to the critical $\kappa_\text{c,OZI}(\alpha)$ obtained before. 
\begin{figure}[htb]
\begin{center}
\scalebox{0.55}{\includeEPSTEX{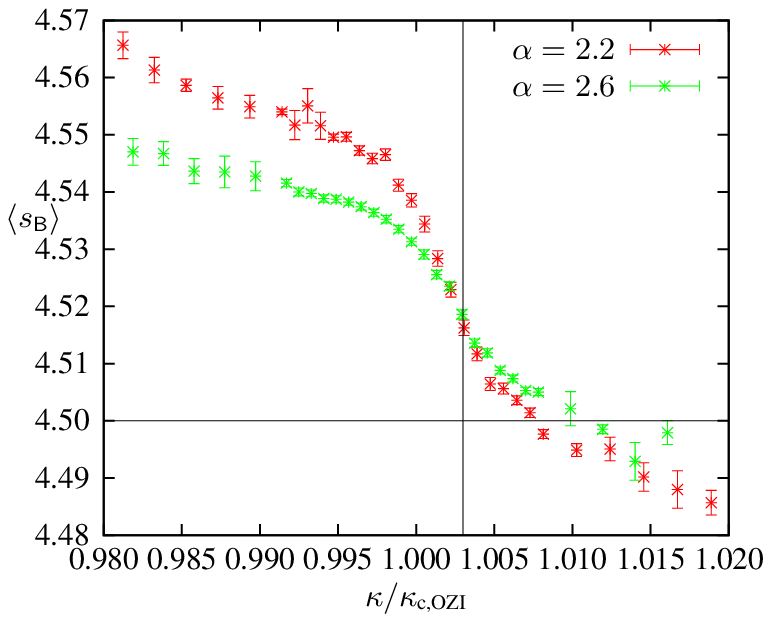}}\hskip5mm
\scalebox{0.55}{\includeEPSTEX{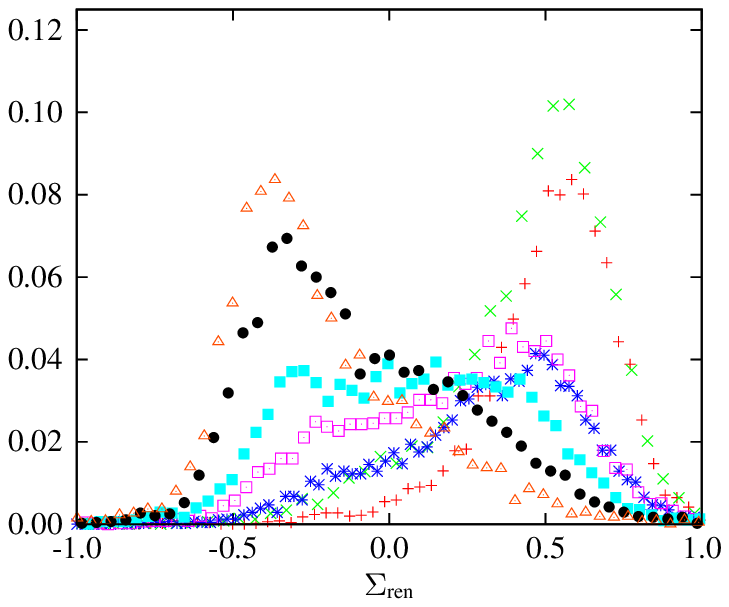}}\hskip5mm
\scalebox{0.55}{\includeEPSTEX{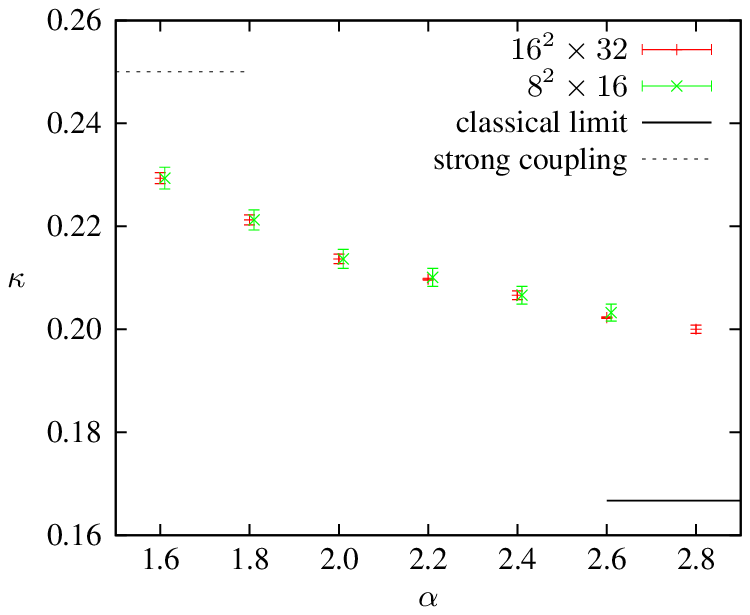}}
\caption{The bosonic action (left panel) is shown for $\alpha=2.2$ and $\alpha=2.6$. The Histograms of the chiral condensate in the center panel point to a first oder phase transition as a function of the renormalized gluino mass. In the right panel the critical line is shown.}
\label{fig:chiralCondensatekappac}
\end{center}
\end{figure}
For both $\alpha=2.2$ and $\alpha=2.6$, the deviations in the critical couplings
are less than $0.5$ percent, see Tab. \ref{tab:kappac}.
\begin{table}[htb]
\begin{center}
\begin{tabular}{c|cc}
$\alpha$ & $\kappa_\text{c,OZI}$ & $\kappa_\text{c}$  \\
\hline\hline $2.2$ & $0.2092(1)$ & $0.2097(4)$ \\
$2.6$ & $0.20167(2)$ & $0.2023(2)$
\end{tabular}
\caption{$\kappa_c$ obtained from the vanishing gluino mass (left) and from the chiral condensate (right).}
\label{tab:kappac}
\end{center}
\end{table}
In the left panel of Fig.~\ref{fig:chiralCondensatekappac}
the bosonic action is plotted. Both curves for $\alpha=2.2$ and $\alpha=2.6$
intersect each other almost exactly at the critical coupling obtained from the
chiral condensate. The deviation from the theoretical value for restoration of
supersymmetry at the intersection point is also about $0.5$ percent, indicating
that a restoration of supersymmetry on the lattice for this model is possible.
As pointed out before, the chiral condensate as a function of $\kappa$ should
undergo a first order phase transition at the point where the gluino mass
vanishes. Therefore, in Fig.~\ref{fig:chiralCondensatekappac} (center panel) histograms of the
chiral condensate for $\alpha=2.2$ are shown in the vicinity of the critical coupling.
The double peak structure and the
coexistence of both phases at the critical coupling clearly point to a first
order transition. This indicates that chiral symmetry is spontaneously
broken in the theory and supersymmetry can be restored in the continuum limit, i.e. the soft breaking of supersymmetry due
to the Wilson mass can be removed by fine-tuning the bare gluino mass. In order
to determine the critical line in the $(\alpha,\kappa)$-plane, the above
sketched analysis is performed for different values of the gauge coupling and on
different lattices. The results for two different lattices $8^3\times 16$
and $16^3 \times 32$ are shown in Fig.~\ref{fig:chiralCondensatekappac} (right panel).

\section{Conclusions and Outlook}

\noindent In the first part we have shown that it is not possible to obtain a 
discretisation for the O(3) NLSM that preserves both O(3) symmetry and part
of supersymmetry on the lattice. Our formulation incorporates 
the O(3) symmetry exactly but breaks supersymmetry. We argued that a 
finetuning of the hopping parameter is sufficient to cancel the explicit 
breaking of chiral symmetry. 

For $\mathcal{N}=2$ Super-Yang-Mills theory in three dimensions we showed explicitly for the gauge
group $SU(2)$ that it is possible to perform a supersymmetric continuum limit by finetuning the theory 
to a vanishing renormalized gluino mass. Our next aim is to investigate the spectrum of bound states as predicted by the effective
lagrangian. However, this is difficult due to the involved disconnected correlation functions which typically
show a very low signal-to-noise ratio.

	


\begin{thebibliography}{99}

\providecommand{\eprint}[1]{ [\href{http://arxiv.org/abs/#1}{arXiv:#1}]}

\bibitem{catterall09}
S.~Catterall, D.~Kaplan and M.~Unsal,
\newblock Phys. Rept. 484 (2009) 71-130.

\bibitem{sigmapaper}
R.~Flore, D.~Körner, A.~Wipf and C. Wozar,
\newblock JHEP 11 (2012) 159.

\bibitem{catterall06}
S.~Catterall and S.~Ghadab,
\newblock JHEP 10 (2006) 063.

\bibitem{kastner08}
T.~Kaestner, G.~Bergner, S.~Uhlmann, A.~Wipf and C.~Wozar,
\newblock Phys.\ Rev.\ D {\bf 78} (2008) 095001.







\bibitem{Veneziano1982231}
G.~Veneziano and S.~Yankielowicz,
\newblock Phys. Lett. B 113 (1982) 231 -- 236.

\bibitem{Ferrara1974239}
S.~Ferrara, B.~Zumino, and J.~Wess,
\newblock Phys. Lett. B 51 (1974) 239 -- 241.

\bibitem{Kirchner:1998mp}
R.~Kirchner, I.~Montvay et al. 
\newblock Phys. Lett. B 446 (1999) 209--215.

\bibitem{Donini:1997hh}
A.~Donini, M.~Guagnelli, P.~Hernandez, and A.~Vladikas,
\newblock Nucl. Phys. B 523 (1998) 529--552.

\bibitem{Curci:1986sm}
G.~Curci and G.~Veneziano,
\newblock Nucl. Phys. B 292 (1987) 555.

\bibitem{Bergner:2011wf}
G.~Bergner, I.~Montvay, G.~Munster, D.~Sandbrink, and U.~D.~Ozugurel,
\newblock \eprint{1111.3012}.

\bibitem{Farchioni:2004fy}
F.~Farchioni and R.~Peetz,
\newblock Eur. Phys. J. C 39 (2005) 87--94.

\bibitem{Giedt:2008xm}
J.~Giedt, R.~Brower, S.~Catterall, G.~Fleming and P.~Vranas,
\newblock Phys. Rev. D 79 (2009) 025015.

\bibitem{Farrar:1997fn}
G.R.~Farrar, G.~Gabadadze, and M.~Schwetz,
\newblock Phys. Rev. D 58 (1998) 015009.

\bibitem{Unsal:2007jx}
Mithat Unsal,
\newblock Phys.Rev. D80:065001, 2009.

\end{thebibliography}
\end{document}